\documentclass[pss]{wiley2sp} % provides pss two-column style
\usepackage{amsmath}

\usepackage{epstopdf}
\usepackage{graphicx}% Include figure files
\usepackage{dcolumn}% Align table columns on decimal point
\usepackage{bm}% bold math
\usepackage{multirow}
\usepackage{booktabs}

\begin{document}

% Title of the article
\title{Comparative Study of Optical and Magneto-Optical Properties of Normal, Disordered and Inverse Spinel Type Oxides}

% Abbreviated title for the page headers
\titlerunning{ Comparative Study of Opt. and Mag.-Opt. Prop. of Normal, Disordered and Inverse Spinel Type Oxides }

% Authors
\author{%
  Vitaly Zviagin\textsuperscript{\Ast,\textsf{\bfseries 1}},
  Peter Richter\textsuperscript{\textsf{\bfseries 2}},
  Tammo B\"ontgen\textsuperscript{\textsf{\bfseries 1}},
  Michael Lorenz\textsuperscript{\textsf{\bfseries 1}},
  Michael Ziese\textsuperscript{\textsf{\bfseries 1}},
  Dietrich R. T. Zahn\textsuperscript{\textsf{\bfseries 2}},
  Georgeta Salvan\textsuperscript{\textsf{\bfseries 2}},
  Marius Grundmann\textsuperscript{\textsf{\bfseries 1}}, and
  R\"udiger Schmidt-Grund\textsuperscript{\textsf{\bfseries 1}}}

% Abbreviated list of authors for the page headers
\authorrunning{V. Zviagin {\it et al.}}

%E-mail-address of corresponding author
\mail{e-mail
  \textsf{vitaly.zviagin@physik.uni-leipzig.de}}

% author's affiliations/addresses
\institute{%
  \textsuperscript{1}\,Universit\"at Leipzig, Institut f\"ur Experimentelle Physik II, Linn\'{e}str. 5, D-04103 Leipzig, Germany\\
  \textsuperscript{2}\,Semiconductor Physics, Technische Universit\"at Chemnitz, D-09107 Chemnitz, Germany\\
  }

\received{XXXX, revised XXXX, accepted XXXX} % do not change, will be filled in by the publisher
\published{XXXX} % do not change, will be filled in by the publisher

% Please select about four verbal keywords for your manuscript.

\abstract{%
% This is a macro for the typesetting of two-column text in an
% abstract. It will typeset the two arguments in \abstcol{}{} as the
% left and right column inside the abstract box. At the
% columnbreak there will be always a columnbreak (\par), so both
% columns start with a new paragraph. No automatic column height
% balancing is done.
%
% If used with a \titlefigure it will silently output both
% parameters as consecutive paragraphs.
%
% The macro is defined exclusively inside the argument of \abstract{};
% if used outside it will raise an error.
%
% Usage: \abstcol{<left column>}{<right column>}
\abstcol{%
  Co$_3$O$_4$, ZnFe$_2$O$_4$, CoFe$_2$O$_4$, ZnCo$_2$O$_4$, and Fe$_3$O$_4$ thin films were fabricated by pulsed laser deposition at high and low temperatures resulting in crystalline single-phase normal, inverse, as well as disordered spinel oxide thin films with smooth surface morphology. The dielectric function, determined by spectroscopic ellipsometry in a wide spectral range from 0.5\,eV to 8.5\,eV, is compared with the magneto-optical response of the dielectric tensor, investigated by magneto-optical Kerr effect (MOKE) spectroscopy in the spectral range from 1.7\,eV to 5.5\,eV with an applied magnetic field of 1.7\,T. Crystal field, inter-valence and inter-sublattice charge transfer transitions, and transitions from O$_{2p}$ to metal cation 3d        }{%
   or 4s bands are identified in both the principal diagonal elements and the magneto-optically active off-diagonal elements of the dielectric tensor. Depending on the degree of cation disorder, resulting in local symmetry distortion, the magneto-optical response is found to be strongest for high crystal quality inverse spinels and for disordered normal spinel structure, contrary to the first principle studies of CoFe$_2$O$_4$ and ZnFe$_2$O$_4$. The results presented provide a basis for deeper understanding of light-matter interaction in this material system that is of vital importance for device-related phenomena and engineering.}}

% The class file requires the standard graphicx Latex package. See the 'LaTeX
% standard graphics and color packages documentation' for more information at
% <http://tug.ctan.org/tex-archive/macros/latex/required/graphics/grfguide.pdf>.
%
% Accepted figure file formats depend on which LaTeX flavour is used.
% Classic LaTeX is always able to use Encapsulated Postscript (EPS);
% PDFLaTeX can't use this but accepts PDF, JPG, PNG, and GIF formats.
%
% See examples for implementing graphics in floating figure environments later in this file.
% If \titlefigure is given, it takes as its mandatory parameter the
% name (without extension) of some figure file.

\maketitle   % please do not remove
\section{Introduction}
Semitransparent spinel oxides continue to receive a considerable amount of attention due to their numerous high-frequency and high-power applications, including microwave devices, magnetic and magneto-optical recording, detectors, sensors, and electronic information mass storage. \cite{Chen,Schein,Timopheev,Venkateshvaran,Raul} With a chemical formula AB$_2$O$_4$, spinel ferrites and cobaltites consist of a cubic close-packed arrangement of oxygen ions, with A$^{2+}$ and B$^{3+}$ ions located at two different crystallographic sites. \cite{Soliman} The ideal crystallization process results in either normal or inverse spinel structure depending on the material, while the growth temperature determines the level of disorder in these structures. \cite{Timopheev,Brachwitz} Co$_3$O$_4$ (CCO), ZnFe$_2$O$_4$ (ZFO), and ZnCo$_2$O$_4$ (ZCO) are ideally crystallizing in a normal spinel structure. They have a general ion distribution formula (A$^{2+}$)[B$^{3+}$]$_2$O$_4^{2-}$, where the divalent A$^{2+}$ ions occupy one eighth of the tetrahedral sites, denoted by parentheses, of the cubic close-packed oxygen lattice, while the trivalent B$^{3+}$ ions occupy one half of the octahedral sites, denoted by square brackets. \cite{Brachwitz} The inverse spinel structure, which is present for ideal CoFe$_2$O$_4$ (CFO) and Fe$_3$O$_4$ (FFO), has a general formula for the ion distribution (B$^{3+}$)[A$^{2+}$B$^{3+}$]O$_4^{2-}$. The tetrahedral sites are occupied by half of $B^{3+}$ cations, while the octahedral sites are occupied by the rest of B$^{3+}$ cations and the A$^{2+}$ cations. \cite{Himcinschi} When A is a magnetic ion, a net magnetic moment of the eight parts forming the unit cell in AB$_2$O$_4$ is present due to uncompensated magnetic moments of the A and/or B magnetic sub-lattices. According to the strength of the oxygen mediated A-O-A or A-O-B coupling materials are either ferrimagnetic or ferromagnetic. \cite{Soliman,Hou,Liskova} \newline \indent
A comparison of the dielectric function of investigated thin films with that of other spinels and similar materials allows an assignment of the transitions. The assigned optical and magneto-optical transitions are related to either inter-valence charge transfer (IVCT), inter-sublattice charge transfer (ISCT) transitions, crystal field (CF) transitions, namely CF splitting of Co$^{2+}$ orbitals, or from O$_{2p}$ band to electronically higher A or B ion bands. Similar electronic structure and electronic transitions have also been found in other transitions metal oxides such as NiFe$_2$O$_4$, CuFe$_2$O$_4$, Li$_{0.5}$Fe$_{2.4}$O$_4$, MgFe$_2$O$_4$. \cite{Himcinschi,Fontijin,Veis,Zhang,Visnovsky}
 Optical investigations of CCO and ZCO have revealed a strong dependence of the electronic structure on the growth method and conditions in relation to the structural properties. \cite{Cook,Wang,Kim2,Kormondy} A smaller lattice constant, narrower spectral features with lower transition energies of the CCO film grown by molecular beam epitaxy (MBE), Ref. 44, as compared to the film grown by pulsed laser deposition (PLD), for example, indicates a slightly higher crystal quality of the MBE film with fewer defects.\cite{Kormondy} However, only few works have been published on the magneto-optical properties of ZCO and CCO.\cite{Ohgushi,Tay} Recent studies have been made on the electronic structure of magnetite \cite{Fonin} as well as theoretical papers on electronic and magnetic properties of ZFO and CFO exhibiting normal, inverse and partially inverse structures.\cite{Soliman,Hou} It was found that highest magnetic moment is expected for a normal spinel unit cell and the lowest for an inverse or disordered spinel structure. \cite{Soliman,Fontijin,Hou,Kim3}
Extensive research of either magneto-optical properties or optical properties of CCO, ZFO, CFO, and FFO has presented experimentally the deviation of the electronic structure from the one expected for the bulk material in both normal and inverse spinel structures. \cite{Fontijin,Liu,Bontgen,Ohgushi,Rai}
It is only recently that optical and magneto-optical properties of ZFO and CFO films were compared to each other and a direct correlation between the lattice structure and the magneto-optical response was investigated.\cite{Himcinschi,Liskova} However, for such a wide series of investigated materials, no such study was performed in dependence on the growth temperature as well as with regard to their electronic structure. Supported by structural analysis, this work presents experimental evidence of lattice distortion, namely presence of Fe$^{3+}$ cations on tetrahedral lattice sites, as an adequate explanation for the observed strength of magneto-optical response, with regard to electronic transitions in a wide spectral range.

\section{Experimental Details}

\subsection{Samples}

ZFO, CFO, ZCO, and CCO thin films were deposited on MgO (100) and FFO on MgAl$_2$O$_4$ (100) substrates by PLD using a KrF excimer laser with the wavelength of 248\,{nm}, a laser pulse energy of 600\,{mJ}, and an energy density on the target of about 2\,{$J cm^{-2}$}. In this study, all films were grown at low temperatures (LT), $\sim$ 300\,{$^{\circ}$C} and high temperatures (HT), $\sim$ 650\,{$^{\circ}$C}, except for FFO, which was grown at HT only. The deposition temperatures were measured at the radiation heater element, but the actual substrate temperature is expected to be lower by up to 80\,{K}.\cite{Bontgen} The film thickness ranges from 120\,{nm} to 450\,{nm}.

\subsection{Structural Properties}

The crystalline structure of the samples was analyzed by X-ray diffraction (XRD) $2\Theta - \omega$ scans using a wide-angle Phillips X'Pert Bragg-Brentano diffractometer and CuK$_{\alpha}$ radiation. The sample surface morphology was examined using a Park systems XE-150 atomic force microscope, operated in dynamic non-contact mode.

\begin{figure}[htb]
\includegraphics[width=1\columnwidth]{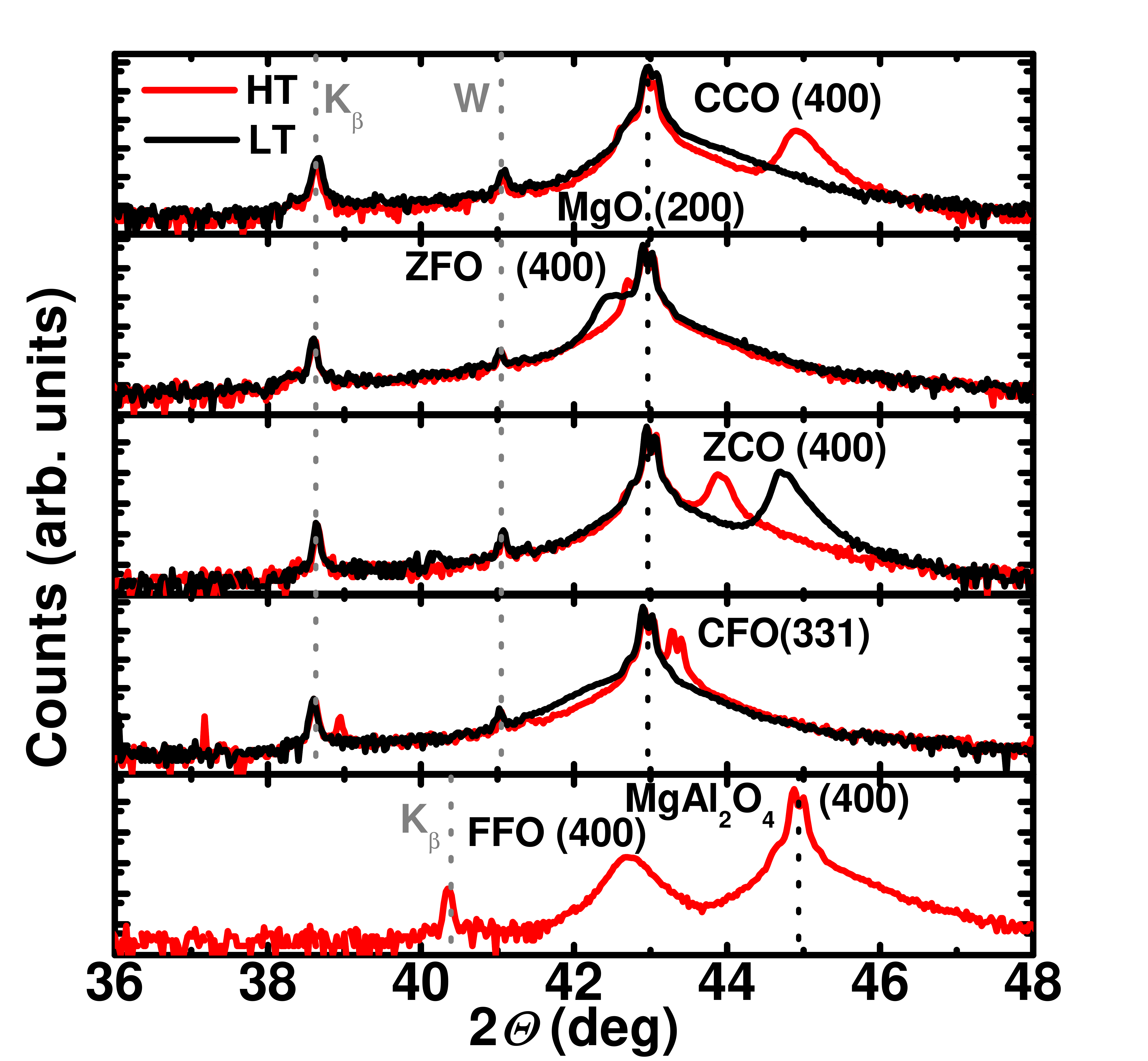}
\sidecaption
\caption{XRD ${2\Theta - \omega}$ scans of the indicated thin films for different growth temperatures (black (LT)/red (HT) lines). The reflexes are indicated. Reflexes marked by $K_{\beta}$ and W are due to the Cu $K_{\beta}$ and W $L_{\alpha}$ spectral lines of the X-ray tube, respectively. The splitting of the substrate reflex is due to the Cu $K_{\alpha1/2}$ splitting.}
\label{Figure1}
\end{figure}

	XRD $2\Theta - \omega$ scans, Fig. \ref{Figure1}, reveal that the thin films contain only single-phase material with no additional observable secondary phases. Parallel orientation of the substrate and crystal axis of the single-phase films, found by XRD $\phi$ scans, show epitaxial relationships (Ref. [7] for films grown at similar conditions). Narrow reflection peaks with $K_{\alpha1}-K_{\alpha2}$ line splitting, visible for the HT CFO thin film, illustrates the high crystalline quality. Due to the inversion of the spinel structure in CFO, a different epitaxial alignment to the substrate, with a (331) orientation is observable.\cite{Kumar} LT CCO and LT CFO films, however, are X-ray amorphous. The lattice constants for ZFO and ZCO films are closer to the bulk values for samples grown at high temperature, as expected for a more ideal cubic structure.\cite{Timopheev,Yao,Spencer,Hanawalt,Paudel} The shift of the lattice constant is due to fewer defects, such as cation disorder and oxygen vacancies, a lattice relaxation with increasing growth temperature is therefore observable.\cite{Lorenz} The lattice constants of thin films, extrapolated from cos$^2(\Theta)$ and the literature bulk values are listed in Table 1.

\begin{figure}[htb]
\includegraphics[width=1\columnwidth]{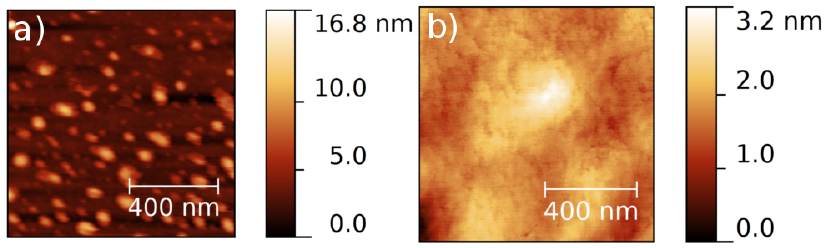}
\caption{Surface morphology of a) LT ZFO and b) HT ZCO with RMS less than 1\,nm measured by AFM.}
\label{Figure2}
\end{figure}

The AFM surface scans reveal a dense and smooth surface morphology for all investigated thin films, shown in Fig. \ref{Figure2} for two extremal examples. Island-like grain structures cover the surface of LT ZFO, CCO, and HT FFO thin films. The root mean square (RMS) value for LT films is larger than for the corresponding HT thin films, except for CCO. The effective medium approximation (EMA) values from ellipsometry modeling (sec. 2.3) are in good agreement with the RMS values.

\begin{table}[b]
\caption{\label{table1}Calculated lattice parameters for each material and bulk values from literature shown for comparison. }
\begin{tabular}[htbp]{ccc}
\hline
Sample&Lattice constant (\AA)&Bulk lattice constant (\AA)\\
\hline
CCO HT & 8.083 & 8.086\cite{Hanawalt}\\
ZFO HT & 8.462 & 8.441\cite{Waerenborgh}\\
ZFO LT & 8.525 & 8.441\cite{Waerenborgh}\\
CFO HT & 8.356 & 8.392 \cite{Hou}\\
ZCO HT & 8.110 & 8.097\cite{Paudel}\\
ZCO LT & 8.256 & 8.097\cite{Paudel}\\
FFO HT & 8.469 & 8.391\cite{Dollase}\\
\hline
\end{tabular}
\end{table}

\subsection{Ellipsometry}
\begin{figure}[htb]
\begin{center}
\includegraphics[width=1\columnwidth]{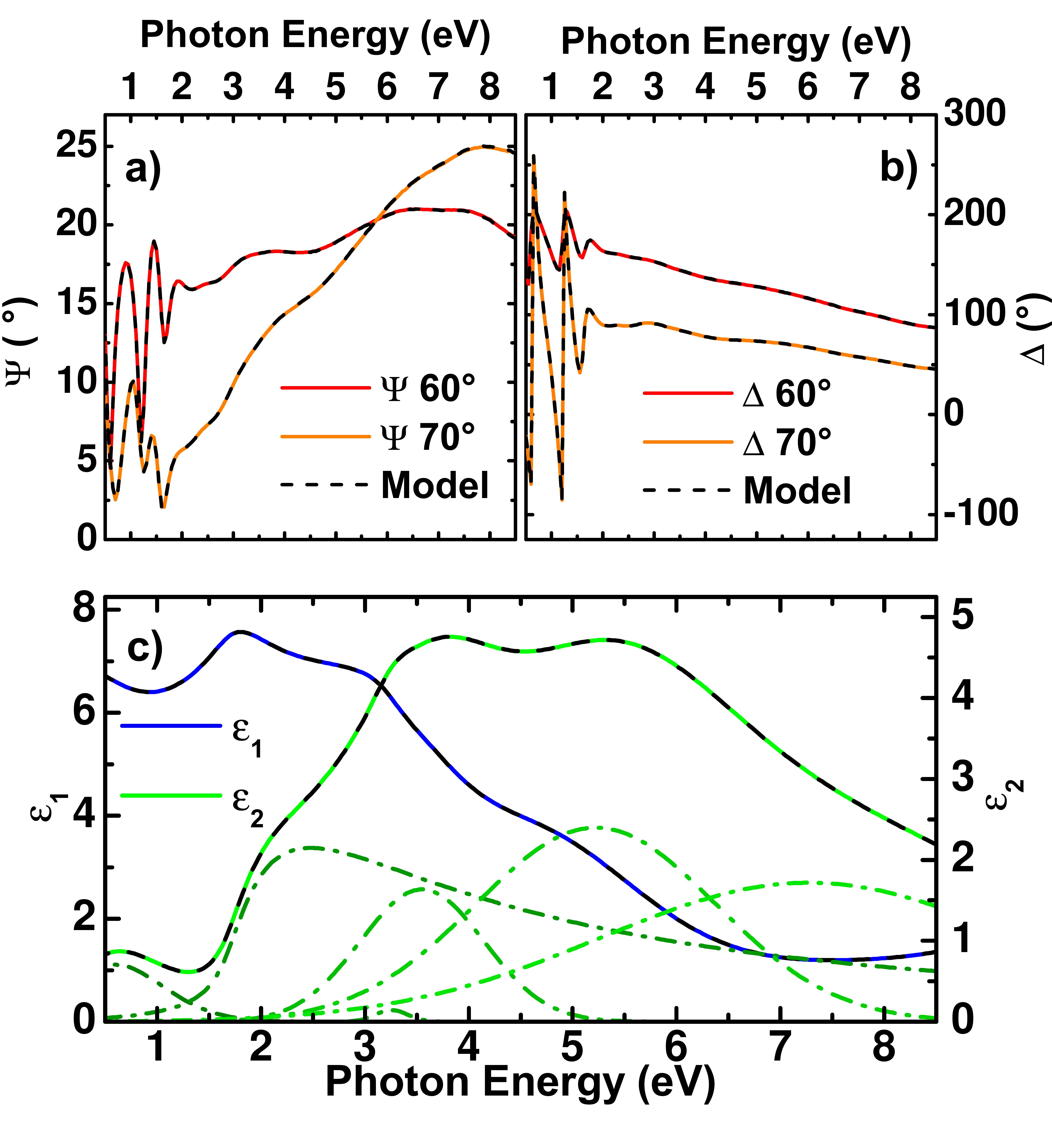}
\caption{a) $\Psi$ and b) $\Delta$ for two different angles of incidence and the respective values calculated from the Kramers Kronig consistent numerical B-Spline MDF approximation (black). c) denoted by the black dashed line is the B-Spline MDF and parametric approximation $\epsilon_1$ (blue) and $\epsilon_2$ (green) for LT CFO film. Green dashed lines show the individual contributions to the parametric model. }
\label{Figure3}
\end{center}
\end{figure}

The dielectric function (DF), $\tilde\varepsilon = \varepsilon_{1} + {\emph {i}}\varepsilon_{2}$ of the thin films was determined by means of spectroscopic ellipsometry in a wide spectral range from 0.5\,eV to 8.5\,eV at room temperature using a {\it J. A. Woollam Co. Inc.} VUV-VASE spectroscopic ellipsometer at angles of incidence of 60\,$^{\circ}$ and 70\,$^{\circ}$. \cite{Fujiwara,Schmidt} For the isotropic case, the measured quantities $\Psi$ and $\Delta$ are defined by the ratio $\rho = r_p/r_s = \tan \Psi \exp (i\Delta)$ of the complex reflection coefficients $r_p$ and $r_s$. The ellipsometry spectra were analyzed by means of a transfer matrix technique for a layer stack model containing a semi-infinite layer for the substrate, a layer for the spinel material, and a surface roughness layer, each described by its thickness and optical constants. The optical constants of each substrate, obtained from modelling the measured experimental data, were found to be in a good agreement with the literature values. \cite{Palik,Zollner,Hosseini} The surface roughness layer is described by a Bruggemann EMA, mixing the DF of the thin film with {50\%} void. Obtained thickness reflects the surface roughness extracted from AFM measurements. \cite{Jellison} Initially the Cauchy approximation for the refractive index $n$ was applied only to HT CFO and ZFO films in the transparent spectral range ($0.5-2.5\,eV$). The optical constants, as well as the thickness, in the entire spectral range were determined by numerically modeling the DF using a Kramers-Kronig consistent B-Spline approximation. \cite{Johs} Based on the obtained optical constants, a parametric dielectric function model was developed for the entire spectral range consisting of a series of up to 8 contributing functions including M0-critical point model functions (CPM), Gaussian and Lorentzian oscillators. \cite{Tiwald} Gaussian-shaped oscillators were used to approximate broad transition features, while Lorenzian-shaped oscillators were used for more narrow features. The CPM functions were used for modeling features at the band-gap energy. \cite{Adachi} Regression analysis was then applied to best match the parametric model dielectric function (MDF) to the numerical MDF. From the numerical MDF, the spectra of the absorption coefficients were obtained. The well suitable model approximation to the experimental data is illustrated for the example of LT CFO, Fig. \ref{Figure3} a) and b). Thickness oscillations below the visible spectral range are observed in the measured spectrum. The corresponding extracted spectra of the real and imaginary parts of the materials dielectric function, obtained from the B-Spline and the parametric approximation including the individual MDF contributions, are shown in Fig. \ref{Figure3} c) exemplary for LT CFO. The parametric MDF for all materials is shown in Fig. \ref{Figure6}.

\subsection{Magneto-Optical Kerr Effect}

The magneto-optical response was measured in the spectral range from 1.7\,eV to 5.5\,eV with an applied field of 1.7\,T in ambient conditions using a home-built magneto-optical Kerr effect (MOKE) setup allowing measurements in polar geometry. \cite{Herrmann} MOKE probes the magneto-optical activity of a sample which can be described by non-zero off-diagonal elements of the dielectric tensor due to the magnetization of the sample. In the performed MOKE experiments, the incoming light is s-polarized and the reflected light is elliptically polarized. The complex Kerr angle, $\Phi_{ \text{Kerr}}$, consisting of $\theta_{\text{Kerr}}$, being the polarization axis rotation, and $\eta_{ \text{Kerr}}$, being ellipticity, can be expressed by Eq.~\ref{eq1}.

\begin{equation}
{\tilde\Phi_{\text{Kerr}} = \theta_{ \text{Kerr}} + {\emph{i}} \cdot {\eta_{ \text{Kerr}} }}.
\label{eq1}
\end{equation}

For semi-transparent samples, interference enhanced MO features are expected due to multi-reflections inside the layers. To account for this effect, a layer thickness dependent modelling procedure for the raw MOKE data was carried out in order to obtain true and pure MO information in the form of the off-diagonal elements of the dielectric tensor.

\begin{figure*}[htb]
\begin{center}
\includegraphics[width=2\columnwidth]{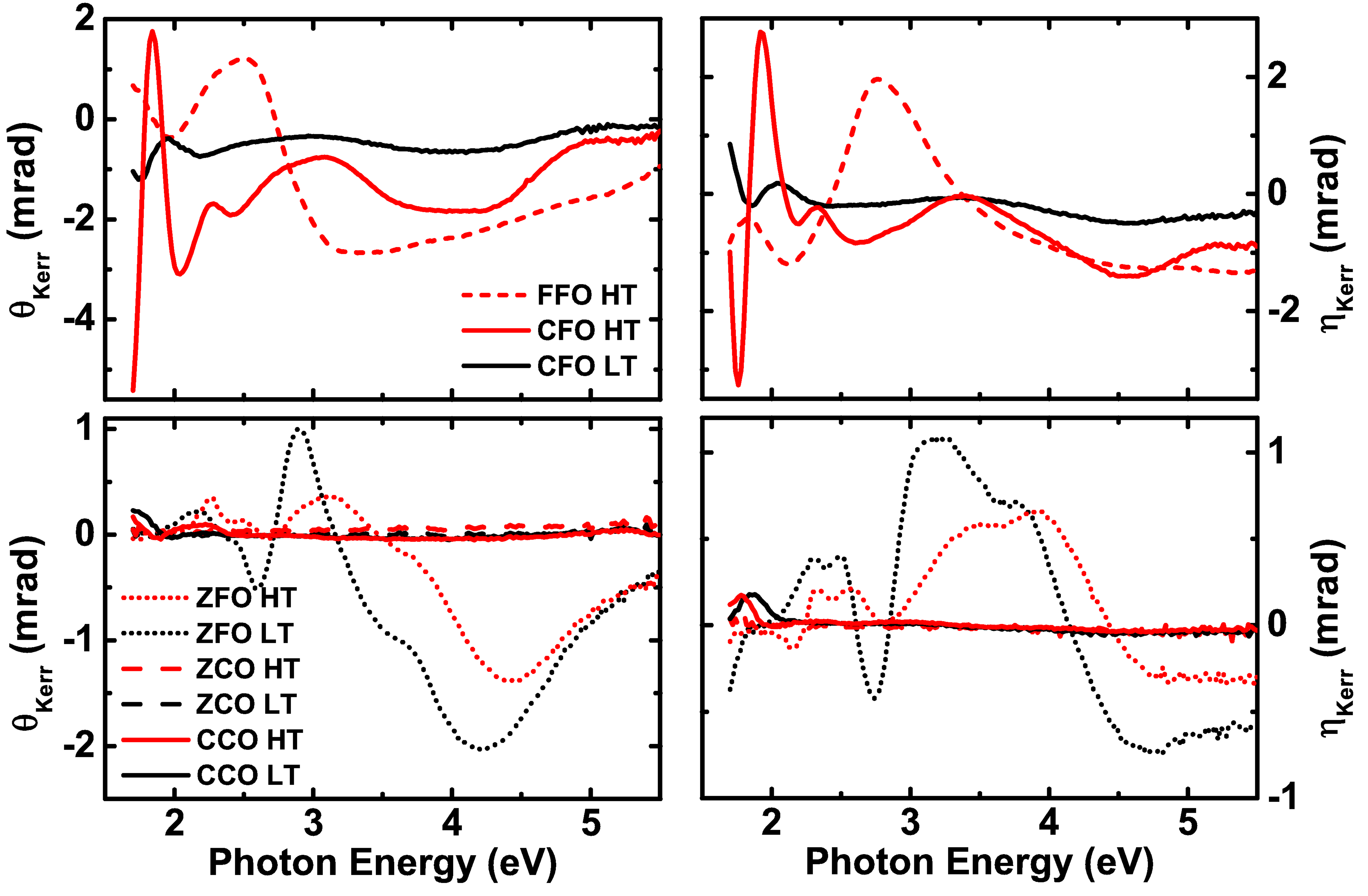}
\caption{ Real, $\theta_{\text{Kerr}}$, (left) and imaginary, $\eta_{\text{Kerr}}$, (right) part of the measured MOKE spectra $\Phi_{\text{Kerr}}$ for FFO, CFO, ZFO, ZCO, and CCO. Note, ZCO and CCO films show little to no magneto-optical response. }
\label{Figure5}
\end{center}
\end{figure*}

In order to obtain $\epsilon_{xy}$ from the $\Phi_{ \text{Kerr} }$ spectra, a parametric modelling procedure involving Gaussian oscillators with positive and negative amplitudes was used to account for the spectral line shape of the experimental MOKE data. This procedure of fitting the MOKE spectra requires also the ellipsometry results as additional input data.\cite{Himcinschi} A detailed theoretical explanation for the model approximation used can be found in literature.\cite{Mack} Only materials containing iron ions were found to show a significant MO response. For CCO, ZCO, and the MgO substrate no MOKE signal was recorded, and none is expected for MgAl$_2$O$_4$. As a result, the off-diagonal elements of the dielectric tensor were obtained for MO active thin films, the corresponding real part, $\epsilon_{1,xy}$, is shown in Fig. \ref{Figure6} right. The sign of $\epsilon_{1,xy}$ is due to the sign convention used in the definition of the complex Kerr rotation angle. The real and imaginary parts of the complex Kerr rotation angle for all samples is shown in Fig. \ref{Figure5}.

\section{Results and Discussion}

\begin{figure*}[htb]
\begin{center}
\includegraphics[width=2\columnwidth]{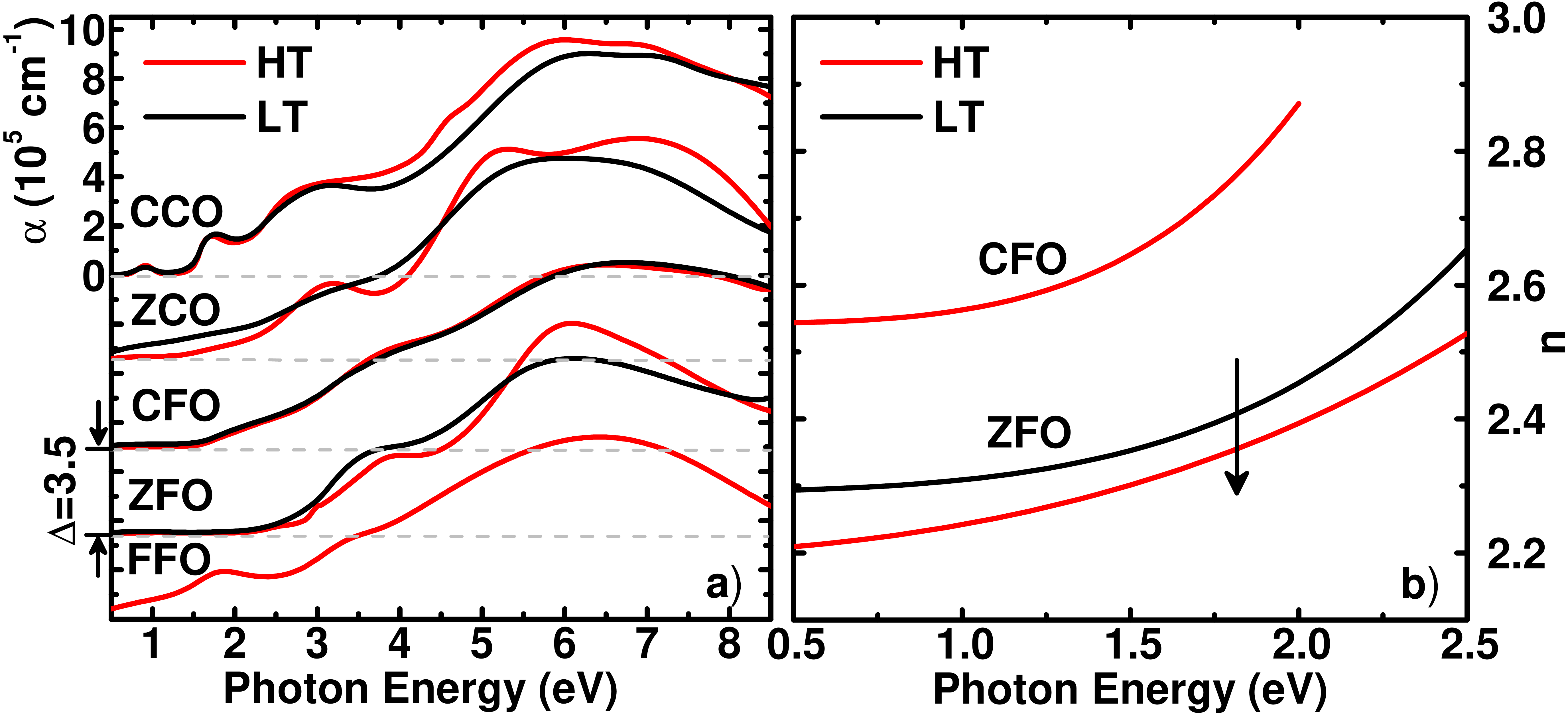}
\caption{ a) Absorption coefficient $\alpha$ derived from the B-Spline MDF. The spectra are shifted vertically against each other for clarity as indicated by gray dashed lines. b) The refractive index in the transparent region was determined from the Cauchy model. }
\label{Figure4}
\end{center}
\end{figure*}

The absorption coefficient spectra, as determined from the numeric B-Spline MDF (sec. 2.3), for all films are dominated by a strong absorption around $\sim6.5\,eV$, while much weaker absorption is present from $\sim1\,eV$ to $\sim3\,eV$ depending on the material as well as the growth temperature (Fig. \ref{Figure4}a)). A blue shift in the absorption edge for binary oxide films with increasing growth temperature is present. \cite{Bontgen} Even though the general line-shape of $\alpha$ does not differ dramatically for different growth temperatures, for ZCO, CCO, and ZFO, the high energy peak shows a strong increase in amplitude with increasing growth temperature. For ZCO the peak additionally splits into two distinct peaks with increasing growth temperature. Only ZFO and HT CFO show a transparent spectral range, for which the refractive index Cauchy model was applied as a model function approximation. A slight decrease of the refractive index with increasing growth temperature for ZFO is visible Fig. \ref{Figure4} b) and consistent with previous works with an estimated uncertainty of $\sim0.03$. \cite{Bontgen} The slight decrease in the refractive index could be caused by variations in the stoichiometry or the film density. It could also be due to different integrated oscillator strength of the electronic transitions at higher energies, following the Kramers-Kronig relation, which changes systematically with increasing growth temperature.\cite{Sareminia,Gaikwad}

\begin{figure*}[htb]
\begin{center}

\includegraphics[width=2\columnwidth]{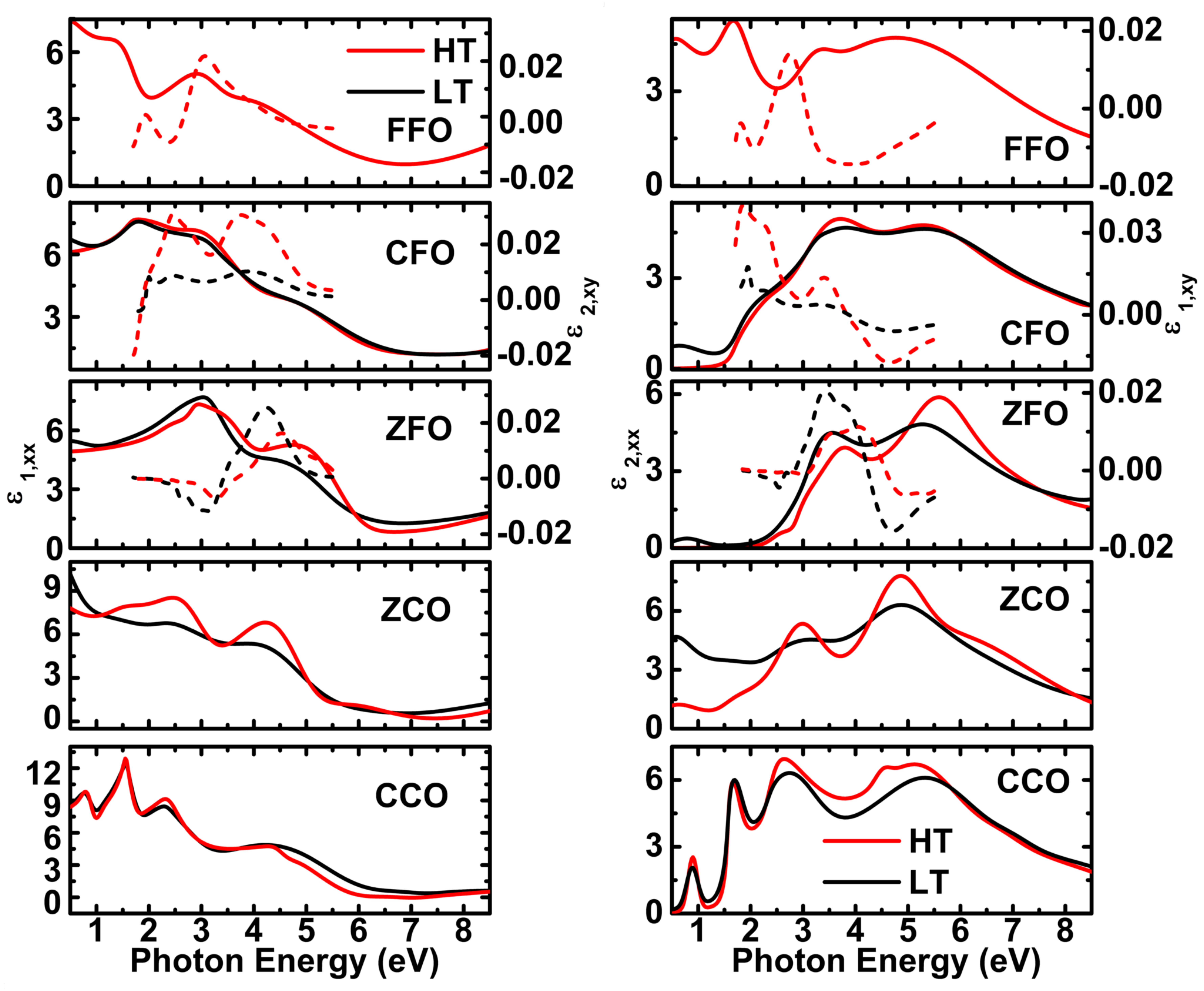}
	\caption{ Imaginary part of diagonal elements $\epsilon_{2,xx}$ (solid lines) and real part of the off-diagonal elements $\epsilon_{1,xy}$ (dashed lines) of the dielectric tensor calculated from the parametric model (right). Spectra of $\epsilon_{1,xx}$ (solid lines) and $\epsilon_{2,xy}$ (dashed lines) (left).}
	\label{Figure6}
\end{center}
\end{figure*}

The modeled spectra show broad and narrow features depending on the growth temperature. At low energies, the features are either very distinctively pronounced or not visible at all, resulting in transparent films in the visible spectral range. For all materials, $\epsilon_{2,xx}$ exhibits a drop in the higher energy range, Fig. \ref{Figure6} right, while $\epsilon_{1,xx}$ shows an increase suggesting additional resonances outside of the measured spectral range, Fig. \ref{Figure6} left, which were accounted for by adding Pole functions to the model.

	The general line-shape of the parametric dielectric function for all materials shows the main dominating features between 5\,eV and 6\,eV, Fig. \ref{Figure6} right, typical for spinel oxides. These features are related to transitions from the O$_{2p}$ band into 4s or higher lying bands of octahedrally coordinated cations, for similar energies reported by B\"ontgen $\emph{et al.}$. \cite{Bontgen} Features above this energy, visible for all samples, apart from FFO, are related to transitions from the O$_{2p}$ band into higher lying bands of tetrahedrally coordinated cations.\cite{Bontgen} The absorption features located below $\sim1\,eV$ are attributed to IVCT transition between octahedrally coordinated Fe$^{2+}$ cations and tetrahedrally coordinated Fe$^{3+}$ cations for FFO and ZFO respectively.\cite{Fontijin,Bontgen,Kim2,Fontijin2} For materials containing Co, the absorption features below 1\,eV correspond to a CF transition within the tetrahedrally coordinated Co$^{2+}$ cation, namely between $^4A_2$ and $^4T_1$ bands, as reported by Kim $\emph{et al.}$ and Fontijn $\emph{et al.}$.\cite{Kim2,Fontijin2} This transition is also assigned to the absorption features visible at $\sim1.9\,eV$ for CFO and ZCO.\cite{Himcinschi,Fontijin2} This transition should not be possible in ZCO, since Co is present only as Co$^{3+}$. For both CCO and ZCO the CF transition is due to direct dipole-forbidden d-d intraband transitions at tetrahedral-site Co$^{2+}$ cations. The transition becomes allowed due to crystal-field splitting in the cubic spinel structure and to hybridization of the O$_{2p}$ states with cobalt 3d states, as argumented by Kormondy $\emph {et al.}$.\cite{Kormondy} We argue that the presence of this peak gives evidence of Co$^{2+}$ in ZCO and thus hints to disorder in the ionic structure. For CCO, however, the absorption feature at $\sim1.7\,eV$ is related to ISCT transition between tetrahedrally coordinated Co$^{3+}$ and octahedrally coordinated Co$^{2+}$ cations, identified by Kim $\emph {et al.}$ and visible in the work by Kormondy $\emph {et al.}$ at the same energy. \cite{Kim2,Kormondy} Consequently, for FFO, the absorption feature at this energy corresponds to an IVCT between octahedrally coordinated Fe$^{2+}$ and Fe$^{3+}$ cations.\cite{Fontijin2,Kim3} While for inverse spinels, FFO and CFO, the absorption features between 2\,eV and 4\,eV can be described by IVCT and ISCT transitions, for normal spinels, ZFO, ZCO and CCO, the features are attributed solely to transitions from the O$_{2p}$ band to bands belonging to different A and B ions of the AB$_2$O$_4$ cubic lattice.\cite{Wemple} The transition energies are in a good agreement with those reported for the same or similar material systems, making the assignment of the transitions accurate. Transition energies and their assignment from literature are listed in Table 2.

\begin{table*}
\caption{\label{tab:table2}Transition energies visible in diagonal, $\epsilon_{2,xx}$, and off-diagonal, $\epsilon_{1,xy}$ for LT and HT films, elements of the dielectric tensor spectra as determined from SE and MOKE, respectively, along with the assignment of the spectral features. The materials, their corresponding spinel structure, and the integrated absolute values of the magneto-optical response over energy are given. The ions situated at octahedral sites are denoted by square brackets and the tetrahedral sites by parentheses.}
\begin{tabular}[htbp]{ccccccccc}
\hline

$AB_2O_4$&Spinel Type&$\int{\left|{\epsilon_{1,xy}}\right|}$& \parbox[t]{1.25cm}{SE $\epsilon_{2,xx}$ energy (eV) \newline LT film}& \parbox[t]{1.25cm}{SE $\epsilon_{2,xx}$ energy (eV) \newline HT film}& \parbox[t]{1.25cm}{MOKE $\epsilon_{1,xy}$ energy (eV)\newline LT film}& \parbox[t]{1.25cm}{MOKE $\epsilon_{1,xy}$ energy (eV) \newline HT film}&\parbox[t]{1.25cm}{Transition type}&\parbox[t]{1.5cm}{Assignment according to literature \newline \cite{Himcinschi,Fontijin,Qiao,Bontgen,Kim2,Fontijin2,Kim3,Wemple}}\\

 \hline

\multirow{7}{*}{FFO HT}&\multirow{7}{*}{Inverse}&\multirow{7}{*}{0.036} &&0.63&&&IVCT& $[Fe^{2+}]t_{2g} \rightarrow [Fe^{2+}]t_{2g}$  \\
&&&&1.76&&2.06&IVCT& $[Fe^{2+}]t_{2g} \rightarrow [Fe^{3+}]e_g$ \\
&&&&&&2.74&ISCT& $(Fe^{3+})t_{2} \rightarrow [Fe^{3+}]t_{2g}$  \\
&&&&&&3.30&IVCT& $[Fe^{2+}]t_{2g} \rightarrow (Fe^{3+})e$  \\
&&&&3.45&&3.51&ISCT& $[Fe^{3+}]e_{g} \rightarrow (Fe^{3+})t_2$  \\
&&&&&&3.92&IVCT& $[Fe^{2+}]t_{2g} \rightarrow (Fe^{3+})t_{2}$  \\
&&&&5.57&&&& $O_{2p} \rightarrow [Fe^{2+}]_{4s}$  \\
\hline

\multirow{5}{*}{CFO HT}&\multirow{5}{*}{Inverse}&\multirow{5}{*}{0.058} &0.66&&&&$CF(Co^{2+})$& $^4A_2 \rightarrow ^4T_1(F)$  \\
&&&&&1.95&1.86&$CF(Co^{2+})$& $^4A_2 \rightarrow ^4T_1(P)$  \\
&&&2.16&2.14&2.26&2.27&IVCT& $[Co^{2+}]t_{2g} \rightarrow [Fe^{3+}]t_{2g}$  \\
&&&&&2.60&2.64&ISCT& $(Fe^{3+})t_{2} \rightarrow [Fe^{3+}]t_{2g}$  \\
&&&&&3.36&3.41&IVCT& $[Co^{2+}]t_{2g} \rightarrow [Fe^{3+}]e_{g}$  \\
\multirow{4}{*}{CFO LT}&\multirow{4}{*}{Disordered}&\multirow{4}{*}{0.019} &3.89&3.87&4.08&3.96&ISCT& $(Fe^{3+})t_{2} \rightarrow [Fe^{3+}]e_{g}$  \\
&&&&&4.72&4.66&& $O_{2p} \rightarrow [Fe^{3+}]_{4s}$  \\
&&&5.97&5.82&&&& $O_{2p} \rightarrow [Fe^{3+}]_{>4s}$  \\
&&&8.47&8.88&&&&$ O_{2p} \rightarrow (Co^{2+})_{>4s}$ \\
\hline

\multirow{4}{*}{ZFO HT}&\multirow{4}{*}{Normal}&\multirow{4}{*}{0.017} &0.80&&&&IVCT& $(Fe^{3 +})t_{2g} \rightarrow (Fe^{3+})e_{g}$ \\
&&&&2.58&2.54&2.52&&$ O_{2p} \rightarrow [Fe^{3+}]_{3d}$  \\
&&&&&3.40&3.58&&$ O_{2p} \rightarrow (Fe^{3+})_{3d}$ \\
&&&3.73&3.88&3.77&4.06&& $O_{2p} \rightarrow [Fe^{3+}]_{3d}$  \\
\multirow{3}{*}{ZFO LT}&\multirow{3}{*}{Disordered}&\multirow{3}{*}{0.033}
&&&4.70&4.94&& $O_{2p} \rightarrow [Fe^{3+}]_{4s}$ \\
&&&5.66&5.91&&5.34&& $O_{2p} \rightarrow [Fe^{3+}]_{>4s}$ \\
&&&8.71&8.80&&&& $O_{2p} \rightarrow (Zn^{2+})_{>4s}$  \\
\hline

\multirow{3}{*}{ZCO HT}&\multirow{3}{*}{Normal}&\multirow{3}{*} &0.68&0.66&&&$CF(Co^{2+})$& $^4A_2 \rightarrow ^4T_1(F) $ \\
&&&1.66&1.82&&&$CF(Co^{2+})$& $^4A_2 \rightarrow ^4T_1(P)$   \\
&&&3.14&3.09&&&&$O_{2p} \rightarrow [Co^{3+}]_{3d}$ \\
\multirow{2}{*}{ZCO LT}&\multirow{2}{*}{Disordered}&\multirow{2}{*} &5.33&5.14&&&&$O_{2p} \rightarrow [Co^{3+}]_{4s}$  \\
&&&6.83&6.69&&&&$ O_{2p} \rightarrow (Zn^{2+})_{>4s}$ \\
\hline

\multirow{3}{*}{CCO HT}&\multirow{3}{*}{Normal}&\multirow{3}{*} &0.90&0.91&&&$CF(Co^{2+})$& $^4A_2 \rightarrow ^4T_1(F) $ \\
&&&1.74&1.71&&&ISCT& $[Co^{3+}]t_{2g} \rightarrow   (Co^{2+})  t_{2g}$   \\
&&&2.94&2.82&&&&$O_{2p} \rightarrow [Co^{3+}]_{3d}$ \\
\multirow{3}{*}{CCO LT}&\multirow{3}{*}{Disordered}&\multirow{3}{*} &&4.65&&&&$O_{2p} \rightarrow (Co^{2+})_{3d}$  \\
&&& 5.35&5.13 &&&& $O_{2p} \rightarrow [Co^{3+}]_{ 3d }$  \\
&&&7.15&6.90&&&& $O_{2p} \rightarrow (Co^{2+})_{>4s}$  \\
\hline

\end{tabular}
\end{table*}

A common feature visible in all MO active materials is located at high energy range, above $\sim4.60\,eV$, and can be interpreted as a transition from the O$_{2p}$ band to 4s and higher lying bands of octahedrally coordinated Fe cations, identified by Fontijn $\emph{et al.}$.\cite{Fontijin} Main features visible in the ZFO spectra are related to transitions between O$_{2p}$ and Fe$^{3+}$ cations. This also includes features located at $\sim2.53\,eV$ and $\sim3.92\,eV$ that are likely to arise from charge transfer transition from the O$_{2p}$ bands to crystal field split t$_{2g}$ and e$_g$ Fe$_{3d}$ bands, similar to transitions observed in iron garnets, reported by Wemple $\emph{et al.}$.\cite{Wemple} Transitions involving octahedrally coordinated Fe$^{3+}$ cation give the energy splitting of approximately 1.39\,eV between its t$_{2g}$ and e$_g$ orbitals, consistent with values reported by Kim $\emph{et al.}$ and Fontijn  $\emph{et al.}$.\cite{Fontijin2,Kim3} The narrow feature dominating the CFO spectra is assigned to the CF transition, located at $\sim1.90\,eV$.\cite{Himcinschi,Fontijin} The other two strong features located at $\sim2.26\,eV$ and $\sim3.38\,eV$ correspond to IVCT transitions between octahedrally coordinated Co$^{2+}$ and Fe$^{3+}$ cations.\cite{Himcinschi,Kim3} As observed by Himcinschi $\emph{et al.}$ and Kim $\emph{et al.}$, a blue shift in the transition energy positions with increasing growth temperature is present for all MO transitions. \cite{Himcinschi,Kim3} Since only a very faint magneto-optical response in the spectral range around the CF transition is visible in CCO and none in ZCO, contrary to expectations, no modelling could be performed for these samples.\cite{Ohgushi}

\section{General Discussion}

	The transitions observed in MOKE and optical analysis were assigned to transitions previously observed for systems of the same materials from literature, which assume a cubic symmetry of both inverse and normal spinel structures.\cite{Himcinschi,Fontijin,Wemple} It should be noted that while the transition energy values are consistent with both Fontjin $\emph{et al.}$ as well as Kim $\emph {et al.}$, the assignment of the mentioned transitions assumes an ideal cubic symmetry with least amount of disorder.\cite{Kim2,Fontijin2} Transitions visible in the optical as well as in the MO spectra provide evidence of either octahedral or tetrahedral symmetry distortion. In particular, features located below $\sim2\,eV$ are stronger for binary and LT ternary oxide films than for the HT ternary oxide films. The considerable strength of the charge transfer transition in normal spinels hints to the degree of inversion in the cationic structure and is supported by first principle studies on the formation of disorder and defects in oxide spinels, as reported by Marcu $\emph{et al.}$ for ZFO.\cite{Marcu} For an inverse spinel structure, however, the inversion symmetry on the octahedral sites can be lost due to distortion of the octahedral symmetry, leading to a stronger charge transfer transition. This transition is most likely to occur in mixed or disordered spinels and shows significant absorption in those that have 2+ and 3+ charged ions in high spin state.\cite{Bontgen}

Moreover, when comparing optical and MO spectra, not all transition features are as pronounced in both of the spectra. Contrary to results reported by Ohgushi $\emph {et al.}$, no MO response due to the CF transition was obtained for CCO, and even for CFO this transition does not show a strong temperature dependence in the MO spectra. Features involving Fe$^{3+}$ cations in both tetrahedral and octahedral sites, however, found in CFO and ZFO, are the dominant features in the MO spectra.\cite{Ohgushi} Their correspondence to the degree of inversion and disorder in both normal and inverse spinel structures provides reasonable explanation for the strength of the MO response of the films. Although Soliman $\emph{et al.}$ and Hou $\emph{et al.}$ have reported that the total magnetic moment of the unit cell is highest for a normal spinel structure and decreases as the degree of inversion increases, from experimentally determined MO active off-diagonal dielectric tensor elements we have found that the MO response is strongest for a high crystalline quality inverse spinel structure.\cite{Soliman,Hou} Future investigations will likely show that the overall magnetic properties of MO active thin films depend strongly on the crystallographic configuration, namely the local tetrahedral and octahedral symmetry and its distortion.

\section{Conclusions}
Normal, disordered, and inverse AB$_2$O$_4$ spinel oxide thin films were deposited by PLD at high and low temperatures on MgO and MgAl$_2$O$_4$ substrates. Smoother surface morphology, decrease in the lattice constant as well as a blue shift in the energies of optical transitions provide direct evidence for the relaxation of the lattice towards the bulk and an increase of the crystal quality with increasing growth temperature. Optical transitions visible in optical and MO spectra are CF, IVCT, ISCT, and transitions from O$_{2p}$ to A or B cation lattice sites. Assignment of these transitions gives an insight into the degree of cation disorder as well as octahedral or tetrahedral symmetry distortion within the lattice. The dominant features in the off-diagonal elements, assigned primarily to transitions involving tetrahedrally and octahedrally coordinated Fe$^{3+}$ cations seen in both CFO and ZFO, provide a reasonable explanation for the observable strength of the MO response. Contrary to first-principles studies, we have found that the MO response is highest for inverse spinel structure and lowest for a normal spinel structure. From this investigation we conclude that high crystalline quality inverse spinel thin films, containing Fe$^{3+}$ cations, are expected to have a stronger MO response with more pronounced features, than a normal spinel structure thin film with distorted local symmetry.

\begin{acknowledgement}
We thank Gabriele Ramm for PLD target preparation and Holger Hochmuth for the thin film growth.
This work was supported by Deutsche Forschungsgemeinschaft within Sonderforschungsbereich 762 - "Functionality of Oxide Interfaces" as well as DFG-Research-Unit 1154 "Towards Molecular Spintronics".
\end{acknowledgement}

\end{document}